\documentclass{JAC2001} 
\usepackage{graphicx}
\newcommand{\ud}{\mathrm{d}}
\begin{document}
\title{ORBITAL BEAM DYNAMICS IN MULTIPOLE FIELDS VIA MULTISCALE EXPANSIONS}
\author{Antonina N. Fedorova,  Michael G. Zeitlin \\ 
IPME, RAS, V.O. Bolshoj pr., 61, 199178, St.~Petersburg, Russia 
\thanks{e-mail: zeitlin@math.ipme.ru}\thanks{ http://www.ipme.ru/zeitlin.html;
http://www.ipme.nw.ru/zeitlin.html}
}

\maketitle

\begin{abstract}
We present the applications of methods from nonlinear local harmonic analysis       
in variational framework to calculations of nonlinear     
motions in polynomial/rational approximations (up to any order) of arbitrary n-pole fields. 
Our approach is based on the methods           
provided possibility to work with dynamical beam/particle localization in      
phase space, which gives        
representions via exact nonlinear high-localized eigenmodes      
expansions and allows to control contribution to motion from each scale     
of underlying multiscale structure.          
\end{abstract}

\section{INTRODUCTION}

In this paper we consider the applications of a new nu\-me\-ri\-cal-analytical 
technique based on the methods of local nonlinear harmonic
analysis or wavelet analysis to the 
calculations of orbital motions in arbitrary n-pole fields.  
Our main examples are motion in
transverse plane for a single
particle in a circular magnetic lattice in case when we take into account
multipolar expansion up to an arbitrary finite number
and particle motion in storage rings.
We reduce initial
dynamical problem to the finite number 
of standard
algebraical problems and represent all dynamical variables as expansion in the
bases of maximally localized in phase space functions.   
Our approach in this paper is based on the generalization 
of variational-wavelet approach from [1]-[12].
Starting in part 2 from Hamiltonians of orbital motion in magnetic lattice
with additional kicks terms and rational approximation of classical motion in storage rings, 
we consider
in part 3 variational-biorthogonal formulation for dynamical system with 
rational nonlinearities
and 
construct
explicit representation for all dynamical variables 
as expansions in nonlinear high-localized eigenmodes.

\section{MOTION IN THE MULTIPOLAR FIELDS}

The magnetic vector potential of a magnet with $2n$ poles in Cartesian
coordinates is
\begin{equation}
A=\sum_n K_nf_n(x,y),
\end{equation}
where $ f_n$ is a homogeneous function of  $x$ and $y$ of order $n$.
The cases $n=2$ to $n=5$
correspond to low-order multipoles: quadrupole, sextupole, octupole, decapole.
The corresponding Hamiltonian is ([13] for designation):
\begin{eqnarray}\label{eq:ham}
&&H(x,p_x,y,p_y,s)=\frac{p_x^2+p_y^2}{2}+\nonumber\\
&&\left(\frac{1}{\rho^2(s)}-k_1(s)\right)
\cdot\frac{x^2}{2}+k_1(s)\frac{y^2}{2}\\
&&-{\cal R}e\left[\sum_{n\geq 2}
\frac{k_n(s)+ij_n(s)}{(n+1)!}\cdot(x+iy)^{(n+1)}\right]\nonumber
\end{eqnarray}
Then we may take into account an arbitrary but finite number of terms in expansion
of RHS of Hamiltonian (\ref{eq:ham}) and
from our point of view the corresponding Hamiltonian equations of motions are
not more than nonlinear ordinary differential equations with polynomial
nonlinearities and variable coefficients.
Also we may add the terms corresponding to kick type contributions 
of rf-cavity:
\begin{eqnarray}
A_\tau=-\frac{L}{2\pi k}\cdot V_0\cdot \cos\big(k\frac{2\pi}
 {L}\tau\big)\cdot\delta(s-s_0)
\end{eqnarray}
or localized cavity $V(s)=V_0\cdot \delta_p(s-s_0)$ with $\delta_p(s-s_0)=
\sum^{n=+\infty}_{n=-\infty}\delta(s-(s_0+n\cdot L))$
at position $s_0$.  
We consider, as the second example, the particle motion in
storage rings in standard approach, which is based on
consideration in [13].
Starting from Hamiltonian, which described classical dynamics in
storage rings
and using Serret--Frenet parametrization, we have after
standard manipulations with truncation of power series expansion of
square root the following
approximated (up to octupoles) Hamiltonian for orbital motion
in machine coordinates:
\begin{eqnarray}
&&{\cal H}=
   \frac{1}{2}\cdot\frac{[p_x+H\cdot z]^2 + [p_z-H\cdot x]^2}
{[1+f(p_\sigma)]}\nonumber\\
&&+p_\sigma-[1+K_x\cdot x+K_z\cdot z]\cdot f(p_\sigma)\\
&&+\frac{1}{2}\cdot[K_x^2+g]\cdot x^2+\frac{1}{2}\cdot[K_z^2-g]\cdot z^2-
  N\cdot xz \nonumber\\
&&+\frac{\lambda}{6}\cdot(x^3-3xz^2)+\frac{\mu}{24}\cdot(z^4-6x^2z^2+x^4)
\nonumber\\
&&+\frac{1}{\beta_0^2}\cdot\frac{L}{2\pi\cdot h}\cdot\frac{eV(s)}{E_0}\cdot
\cos\left[h\cdot\frac{2\pi}{L}\cdot\sigma+\varphi\right]\nonumber
\end{eqnarray}
Then we use series expansion of function $f(p_\sigma)$ from [13]:
$
f(p_\sigma)=f(0)+f^\prime(0)p_\sigma+f^{\prime\prime}(0)
p_\sigma^2/2+\ldots
=p_\sigma-
p_\sigma^2/(2\gamma_0^2)+\ldots$
and the corresponding expansion of RHS of equations corresponding to (4).
In the following we take into account only  arbitrary
polynomial/rational  (in terms of dynamical variables) expressions.

\section{VARIATIONAL APPROACH IN BIORTHO\-GONAL WAVELET BASES}

The first main part of our consideration is some variational approach
to these problems, which reduce initial problem to the problem of
solution of functional equations at the first stage and some
algebraical problems at the second stage.
Multiresolution expansion is the second main part of our construction.
As a result the solution is parameterized by solutions of two reduced algebraical
problems, one is nonlinear and others are linear
problems obtained from wavelet
constructions and represented as expansion
in a compactly supported wavelet basis.
Because integrand of variational functionals is represented
by bilinear form (scalar product) it seems more reasonable to
consider wavelet constructions which take into account all advantages of
this structure.
Let $(M,\omega)$ be a symplectic
manifold, $H$ is Hamiltonian, $X_H$ is
unique Hamiltonian vector field defined  by
$
\omega(X_H(x),\upsilon)=-dH(x)(\upsilon),\quad \upsilon\in T_xM,
\quad x\in M,
$
where $ \omega$ is the symplectic structure.
T-periodic solution $x(t)$ of the Hamiltonian equations
$
\dot x=X_H(x)
$ on M
is a solution, satisfying the boundary conditions $x(T)$ $=x(0), T>0$.
Let us define a function 
\begin{equation}
\Phi(x)=\displaystyle\int_0^T\frac{1}{2}<-J\dot x, x>dt-
\int_0^T H(x(t))dt
\end{equation}
The critical points of $\Phi$ are the periodic solutions of $\dot x=X_H(x)$.
Computing the derivative at $x\in\Omega$ in the direction of $y\in\Omega$,
we find
\begin{eqnarray}
&&\Phi'(x)(y)=\frac{d}{d\epsilon}\Phi(x+\epsilon y)\vert_{\epsilon=0}\\
&&=
\displaystyle\int_0^T<-J\dot x-\bigtriangledown H(x),y>dt\nonumber
\end{eqnarray}
Consequently, $\Phi'(x)(y)=0$ for all $y\in\Omega$ iff the loop $x$ satisfies
the equation
\begin{equation}
-J\dot x(t)-\bigtriangledown H(x(t))=0
\end{equation}
Now we
introduce wavelets in our underlying bilinear structure.
Let us consider action of operator $S$ on $x$
\begin{equation}
S(\omega (J),H,x,\partial/\partial t,\nabla,t)x=-J\dot x(t)-\bigtriangledown H(x(t))
\end{equation}
which is polynomial/rational in $x$, and have arbitrary dependence on $t$. Then (6) is equivalent to 
\begin{equation}
\int<Sx,y>\ud t=0
\end{equation}
We start with two hierarchical sequences of approximations spaces [14]:
\begin{eqnarray}
&&\dots V_{-2}\subset V_{-1}\subset V_{0}\subset V_{1}\subset V_{2}\dots,\\
&&\dots \widetilde{V}_{-2}\subset\widetilde{V}_{-1}\subset
\widetilde{V}_{0}\subset\widetilde{V}_{1}\subset\widetilde{V}_{2}\dots,\nonumber
\end{eqnarray}
and corresponding biorthogonal expansions:
\begin{eqnarray}
x^N(t)=\sum^N_{r=1}a_r\psi_r(t), \quad
y^N(t)=\sum^N_{k=1}b_k\tilde{\psi}_k(t)
\end{eqnarray}
Let
$W_0$ be complement to $V_0$ in $V_1$, but not necessarily orthogonal
complement.
Orthogonality conditions have the following form:
$\widetilde {W}_{0}\perp V_0$, $\ W_{0}\perp\widetilde{V}_{0}$, $\
V_j\perp\widetilde{W}_j$, $\ \widetilde{V}_j\perp W_j$.
Translates of $\psi$ $\mathrm{span}$ $ W_0$,
translates of $\tilde\psi \quad \mathrm{span} \quad\widetilde{W}_0$.
Biorthogonality conditions are
\begin{equation}
<\psi_{jk},\tilde{\psi}_{j'k'}>=
\int^\infty_{-\infty}\psi_{jk}(x)\tilde\psi_{j'k'}(x)\ud x=
\delta_{kk'}\delta_{jj'},
\end{equation}
 where
$\psi_{jk}(x)=2^{j/2}\psi(2^jx-k)$.
Functions $\varphi(x), \tilde\varphi(x-k)$ form  dual pair:
$
<\varphi(x-k),\tilde\varphi(x-\ell)>=\delta_{kl}$,
 $<\varphi(x-k),\tilde\psi(x-\ell)>=0$.
Functions $\varphi, \tilde\varphi$ generate a multiresolution analysis.
$\varphi(x-k)$, $\psi(x-k)$ are synthesis functions,
$\tilde\varphi(x-\ell)$, $\tilde\psi(x-\ell)$ are analysis functions.
Synthesis functions are biorthogonal to analysis functions. Scaling spaces
are orthogonal to dual wavelet spaces.
Two multiresolutions are intertwining
$
V_j+W_j=V_{j+1}, \quad \widetilde V_j+ \widetilde W_j = \widetilde V_{j+1}
$.
These are direct sums but not orthogonal sums.
So, our representation (11) for solution on the level of resolution  $V_j$ has now the form
\begin{equation}
x_j(t)=\sum_{k}\tilde a_{jk}\psi_{jk}(t),
\end{equation}
where synthesis wavelets are used to synthesize the function. But
$\tilde a_{jk}$ come from inner products with analysis wavelets.
Biorthogonality yields
\begin{equation}
\tilde a_{jm}=\int x^j(t)\tilde{\psi}_{jm}(t) \ud t.
\end{equation}
So, we may use this more useful construction in
our variational approach [1]-[12]. We have modification only on the level of
computing coefficients of reduced nonlinear algebraical system of equations.
This biorthogonal construction is more flexible 
and stable under the action of large
class of operators while orthogonal (one scale for multiresolution)
is fragile, all computations are much more simpler and we accelerate
the rate of convergence. In all types of Hamiltonian calculation,
which are based on some bilinear structures (symplectic or
Poissonian structures, bilinear form of integrand in variational
integral) this framework leads to greater success.
In numerical modelling we may consider very useful wavelet packets.
As a result
we have from (9) the following reduced system of algebraical equations (RSAE)
on the set of unknown coefficients $a_i$ of
expansions (11):
\begin{eqnarray}\label{eq:pol2}
L(S_{ij},a,\alpha_I,\beta_J)=0
\end{eqnarray}
where operator L is algebraization of initial problem 
(8).
$I=(i_1,...,i_{q+2}), \ J=(j_1,...,j_{p+1})$ are multiindexes, by which are 
labelled $\alpha_I$ and $\beta_I$, the  other coefficients of RSAE (\ref{eq:pol2}):
\begin{equation}\label{eq:beta}
\beta_J=\{\beta_{j_1...j_{p+1}}\}=\int\prod_{1\leq j_k\leq p+1}\varphi_{j_k},
\end{equation}
\begin{eqnarray}\label{eq:alpha}
\alpha_I=\{\alpha_{i_1}...\alpha_{i_{q+2}}\}=\sum_{i_1,...,i_{q+2}}\int
\varphi_{i_1}...\dot{\varphi_{i_s}}...\varphi_{i_{q+2}},\nonumber
\end{eqnarray}
where p (q) is the degree of nominator (denominator) part of  operator $S$ (8), 
$i_\ell=(1,...,q+2)$, $\dot{\varphi_{i_s}}=\ud\varphi_{i_s}/\ud t$.
Now, when we solve RSAE (\ref{eq:pol2}) and determine
unknown coefficients from formal expansion (11) we therefore
obtain the solution of our initial problem.
It should be noted that if we consider only truncated expansion with N terms
then we have from (\ref{eq:pol2}) the system of $N\times n$ ($n$ is dimension of $x$ (7))
algebraical equations 
and the degree of this algebraical system coincides
with degree of initial differential system.
The problem of
computations of coefficients $\alpha_I$, $\beta_J$ (\ref{eq:beta})
of reduced algebraical
system may be explicitly solved in wavelet approach.
The bases  functions $\psi_k(t)$ (11)
are obtained via multiresolution expansions (10) and represented by
compactly supported wavelets.
Because affine
group of translations and dilations is inside the approach, this
method resembles the action of a microscope. We have contribution to
final result from each scale of resolution from the whole
infinite scale of spaces (10). 
The solution has the following form
\begin{equation}\label{eq:z}
x(t)=x_N^{slow}(t)+\sum_{j\geq N}x_j(\omega_jt), \quad \omega_j\sim 2^j
\end{equation}
\begin{figure}[htb]
\centering
\includegraphics*[width=60mm]{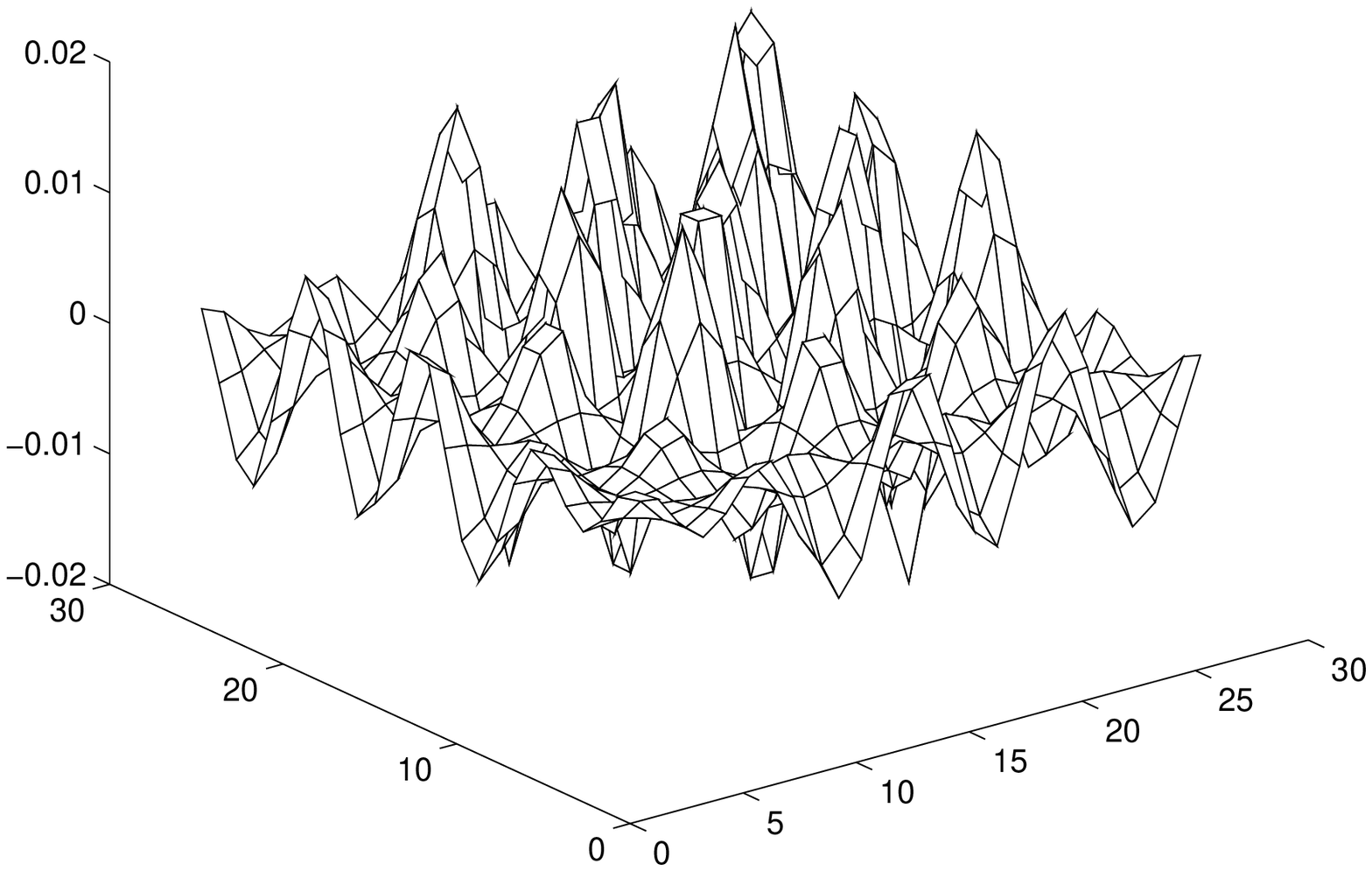}
\includegraphics*[width=60mm]{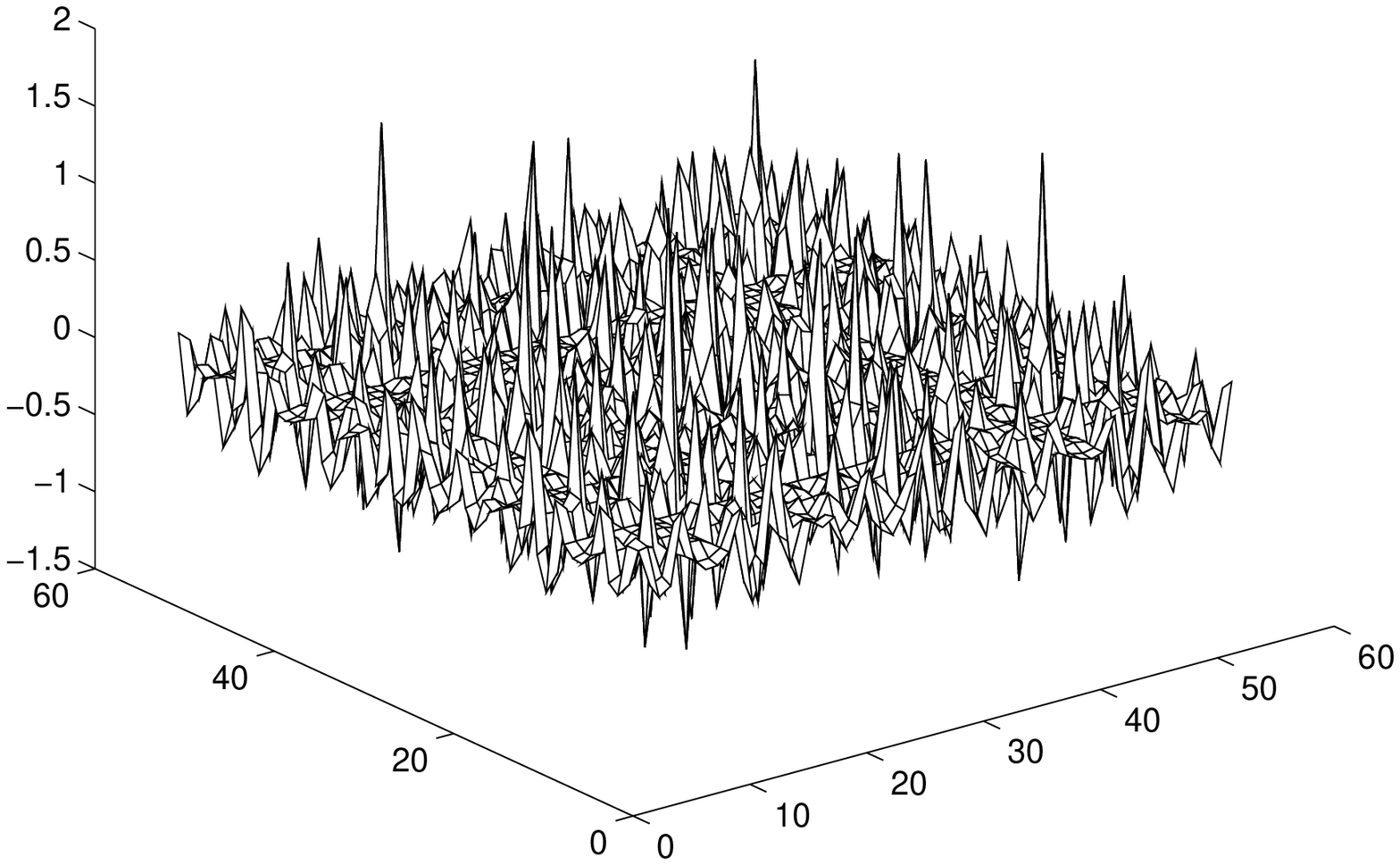}
\caption{Multiscale representations for $x-p_x$  sections.}
\end{figure}
which corresponds to the full multiresolution expansion in all time 
scales.
Formula (\ref{eq:z}) gives us expansion into a slow part $z_N^{slow}$
and fast oscillating parts for arbitrary N. So, we may move
from coarse scales of resolution to the 
finest one for obtaining more detailed information about our dynamical process.
The first term in the RHS of equation (19) corresponds on the global level
of function space decomposition to  resolution space and the second one
to detail space. In this way we give contribution to our full solution
from each scale of resolution or each time scale.
On Fig.~1  we present sections $x-p_x$ corresponding to model (5) in different parameter
regions.

\section{ACKNOWLEDGMENTS}

We would like to thank The U.S. Civilian Research \& Development Foundation (CRDF) for
support (Grants TGP-454, 455), which gave us the possibility to present our nine papers during
PAC2001 Conference in Chicago and Ms. Camille de Walder from CRDF for her help and encouragement.


\begin{thebibliography}{13}

\bibitem{1}
A.N. Fedorova and M.G. Zeitlin, 
 {\it Math. and Comp. in Simulation}, {\bf 46}, 527, 1998.

\bibitem{2}
A.N. Fedorova and M.G. Zeitlin,
{\it New Applications of Nonlinear and Chaotic Dynamics in Mechanics}, 31, 101
Klu\-wer,  1998.

\bibitem{3}
A.N. Fedorova and M.G. Zeitlin,
{\bf CP405}, 87, American Institute of Physics, 1997.\\
Los Alamos preprint, physics/9710035.

\bibitem{4}
A.N. Fedorova, M.G. Zeitlin and Z.~Parsa, 
Proc. PAC97 
{\bf 2}, 1502, 1505, 1508, APS/IEEE, 1998.

\bibitem{5}
A.N. Fedorova, M.G. Zeitlin and Z.~Parsa, 
Proc. EPAC98, 930, 933, Institute of Physics, 1998.

\bibitem{6}
A.N. Fedorova, M.G. Zeitlin and Z.~Parsa,    
{\bf CP468}, 48, American Institute of Physics, 1999.
Los Alamos preprint, physics/990262.

\bibitem{7}
A.N. Fedorova, M.G. Zeitlin and Z.~Parsa,  
{\bf CP468}, 69, American Institute of Physics, 1999.
Los Alamos preprint, physics/990263.

\bibitem{8}
A.N. Fedorova and M.G. Zeitlin,  
Proc. PAC99, 
1614, 1617, 1620, 2900, 2903,
2906, 2909, 2912, APS/IEEE, New York, 1999.
Los Alamos preprints: 
physics/9904039,\\ physics/9904040,
 physics/9904041, physics/9904042,\\
 physics/9904043, 
physics/9904045, physics/9904046,\\
 physics/9904047.

\bibitem{9}
A.N. Fedorova and M.G. Zeitlin,
The Physics of High Brightness Beams, 235, World Scientific, 2000. 
Los Alamos preprint: physics/0003095.

\bibitem{10}
A.N. Fedorova and M.G. Zeitlin,  Proc. EPAC00, 415, 872,  1101, 1190, 1339, 2325,Austrian Acad.Sci.,2000.\\ 
Los Alamos preprints: physics/0008045, physics/0008046,\\
 physics/0008047, physics/0008048, physics/0008049,\\
 physics/0008050.

\bibitem{11}
A.N. Fedorova, M.G. Zeitlin, Proc. 20 International Linac Conf., 300, 303, SLAC, Stanford, 2000. 
Los Alamos pre\-pri\-nts: physics/0008043, physics/0008200.

\bibitem{12}
A.N. Fedorova, M.G. Zeitlin, Los Alamos preprints:\\ physics/0101006, physics/0101007
and World Scientific, in press. 

\bibitem{13}
A. Dragt, Lectures on Nonlinear Dynamics, 1996.\\
A. Bazzarini, e.a., CERN 94-02.

\bibitem{14}

A. Cohen, e.a., CPAM, 45, 485, 1992
\end{thebibliography}
\end{document}